\documentclass[runningheads]{llncs}

\usepackage{graphicx}
\usepackage{amsmath, amssymb}
\usepackage{tabularx}
\usepackage{booktabs}
\usepackage{float}
\usepackage[table]{xcolor} 
\newcolumntype{C}{>{\centering\arraybackslash}X}

\begin{document}

\title{EAGLE: An Efficient Global Attention Lesion Segmentation Model for Hepatic Echinococcosis}

\titlerunning{EAGLE: Lesion Segmentation for HE}

\author{Jiayan Chen \inst{1,3} \and Kai Li \inst{2} \and Yulu Zhao \inst{4} \and Jianqiang Huang \inst{1,3} \and Zhan Wang\inst{5,6}} 
\authorrunning{Chen et al.}

\institute{School of Computer Technology and Application, Qinghai University
\and
Department of Computer Science and Technology, Tsinghua University
\and
Intelligent Computing and Application Laboratory of Qinghai Province, Qinghai University
\and
School of Chemical Engineering, Qinghai University
\and
Department of General Surgery, Affiliated Hospital of Qinghai University
\and
Department of Medical Engineering Integration and Translational Application, Affiliated Hospital of Qinghai University
}

\maketitle              

\begin{abstract}

Hepatic echinococcosis (HE) is a widespread parasitic disease in underdeveloped pastoral areas with limited medical resources. While CNN-based and Transformer-based models have been widely applied to medical image segmentation, CNNs lack global context modeling due to local receptive fields, and Transformers, though capable of capturing long-range dependencies, are computationally expensive. Recently, state space models (SSMs), such as Mamba, have gained attention for their ability to model long sequences with linear complexity. In this paper, we propose EAGLE, a U-shaped network composed of a Progressive Visual State Space (PVSS) encoder and a Hybrid Visual State Space (HVSS) decoder that work collaboratively to achieve efficient and accurate segmentation of hepatic echinococcosis (HE) lesions. The proposed Convolutional Vision State Space Block (CVSSB) module is designed to fuse local and global features, while the Haar Wavelet Transformation Block (HWTB) module compresses spatial information into the channel dimension to enable lossless downsampling. Due to the lack of publicly available HE datasets, we collected CT slices from 260 patients at a local hospital. Experimental results show that EAGLE achieves state-of-the-art performance with a Dice Similarity Coefficient (DSC) of 89.76\%, surpassing MSVM-UNet by 1.61\%.

\keywords{hepatic echinococcosis \and medical image segmentation \and state-space models \and deep learning}
\end{abstract}

\section{Introduction}

Hepatic Echinococcosis (HE) is a zoonotic disease caused by tapeworms of the Echinococcus genus\cite{mcmanus2003echinococcosis,edam}, primarily including cystic echinococcosis (CE) and alveolar echinococcosis (AE)\cite{deplazes2017global,torgerson2015world}. HE is endemic in many regions with extensive animal husbandry but underdeveloped healthcare infrastructure, including western China, East Africa, Central Asia, and South America\cite{torgerson2010global}. The disease burden is particularly high in rural and resource-limited areas, where access to timely diagnosis and treatment remains a challenge. Computed Tomography (CT) is one of the most widely used diagnostic tools for HE. However, the lesions often exhibit complex imaging characteristics, such as indistinct boundaries, varying shapes and sizes, and densities similar to adjacent tissues, making accurate lesion identification particularly difficult.

In recent years, with the rapid development of deep learning, convolutional neural networks (CNNs)\cite{cnn,li2024subnetwork}, especially U-Net\cite{unet}, have been widely applied in medical image segmentation. ResUNet\cite{resunet} introduces residual connections to ease gradient degradation, while UNet++\cite{zhou2018unet++} employs nested dense skip pathways to narrow the semantic gap between encoder and decoder features. However, the inherent locality of convolution operations and the loss of spatial details during downsampling limit the ability of CNN-based models to capture global context and fine-grained features, thus affecting overall segmentation performance—particularly in challenging tasks such as HE lesion segmentation. 
Transformer\cite{transformer} initially achieved great success in the field of natural language processing (NLP) by leveraging multi-head self-attention (MHSA) mechanisms to effectively model long-range dependencies \cite{li2022efficient,xu2025tiger,li2025apollo}. Methods such as Swin Transformer\cite{swintransformer} and TransUNet\cite{transunet} integrate CNNs with Transformers to capture both local and global features, thereby enhancing segmentation performance. However, the computational cost of self-attention grows quadratically with image resolution, and Transformer-based models typically require large-scale pretraining datasets. In domains like HE, where annotated data are scarce and costly to obtain, such requirements pose significant challenges to their practical application.

State Space Models (SSMs) have gained significant attention in recent years due to their efficiency in long-sequence modeling and linear time complexity. Mamba\cite{mamba} introduces a gated state update mechanism and efficient state space convolution, demonstrating superior performance over Transformers across various sequence modeling tasks \cite{li2024spmamba}. Its variant, VMamba\cite{vmamba}, extends Mamba to visual tasks by proposing the SS2D module, enabling efficient processing of two-dimensional image data. Swin-UMamba\cite{swinumamba} and HMT-UNet\cite{hmtunet} incorporate a hybrid architecture that integrates CNNs and SSMs. However, it lacks a dedicated multi-scale feature fusion mechanism, which limits its performance when dealing with HE lesions that exhibit diverse morphological variations. Although MSVM-UNet\cite{msvmunet,li2024audio} captures multi-scale representations in the encoding stage, the absence of corresponding mechanisms in the decoding stage also results in suboptimal performance.

To address these challenges and alleviate the burden on medical resources in remote pastoral areas, we propose EAGLE, an efficient global attention lesion segmentation model for HE. This model comprises two main components: a Progressive Visual State Space (PVSS) Encoder and a Hybrid Visual State Space (HVSS) Decoder.

The main contributions of this work can be summarized as follows:
\begin{enumerate}
\item	We propose EAGLE, a novel model specifically designed for HE lesion segmentation. This model aims to assist radiologists, especially in resource-limited areas, by improving the efficiency and accuracy of HE diagnosis.

\item We introduce two key components: the Convolutional Vision State Space Block (CVSSB) and the Haar Wavelet Transform Block (HWTB). The former integrates convolution and state space modeling through a Depthwise-Aware FeedForward Network (DA-FFN), and is applied in both the encoder and decoder to enhance feature representation. The latter compresses spatial features into the channel dimension during the encoding stage, enabling lossless downsampling.
\item	We conduct extensive experiments on a self-collected dataset containing CT images from 260 HE patients. The results demonstrate that EAGLE achieves state-of-the-art performance, with a DSC of 89.76\%, highlighting its effectiveness in HE lesion segmentation.
\end{enumerate}

\section{Methods}

\subsection{Overall Architecture}

\begin{figure}[h]
\centering
\includegraphics[width=\linewidth]{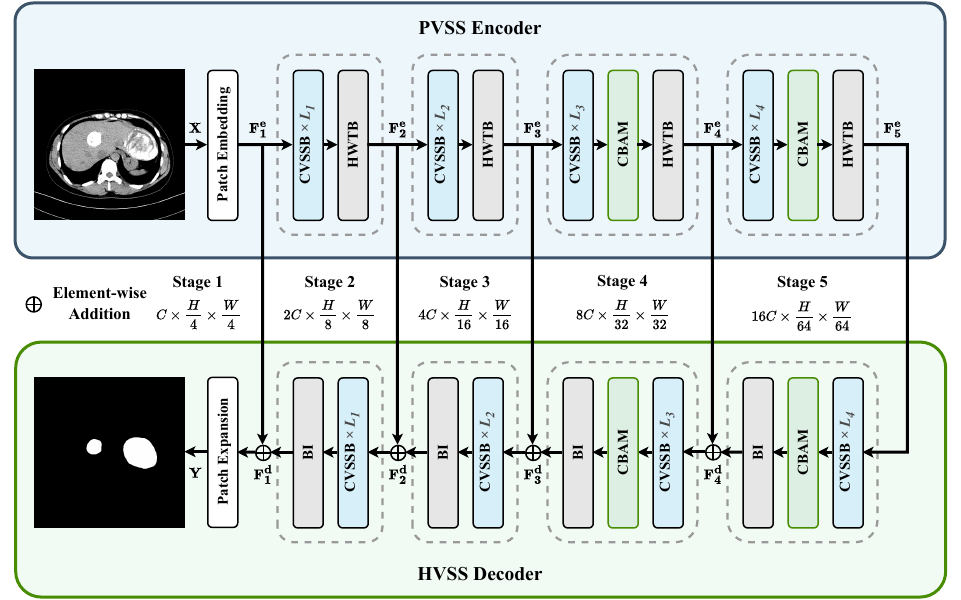}
\vspace{-1em}
\caption{The overall pipeline of EAGLE, which is composed of three main components: a Progressive Visual State Space (PVSS) Encoder, a Hybrid Visual State Space (HVSS) Decoder, and skip connections. $\mathbf{F_i^e}$ denotes the feature map output by the encoder block at the i-th stage, and  $\mathbf{F_i^d}$ denotes the feature map output by the decoder block at the i-th stage.} 
\label{fig:overall}
\end{figure}

Given a CT image $\mathbf{X} \in \mathbb{R} ^{C \times H \times W} $, where $C$, $H$, and $W$ denote the number of channels, height, and width of the image, respectively. With the input being a grayscale image, $C = 1$. Specifically, as show in Figure.~\ref{fig:overall}, the overall pipeline of EAGLE can be summarized as follows:
First, the input image is fed into a patch embedding layer, which partitions the input $\mathbf{X}$ into non-overlapping $4 \times 4$ patches and projects them into a feature map with $C$ channels, denoted as $\mathbf{F}^e_1$. This is followed by four hierarchical encoder blocks, each composed of a stack of CVSSB blocks and a HWTB, which progressively extract feature representations at different scales $\{\mathbf{F}_i^e \in \mathbb{R}^{(32\times 2^{i-1}) \times \frac{H}{2^{i+1}} \times \frac{W}{2^{i+1}}} \mid i \in [2,5]\}$.
In stages 4 and 5, the feature maps are passed through the CBAM\cite{cbam} module after CVSSB to perform global feature weighting. The encoded feature $\mathbf{F}^e_5$ is fed into the decoder to generate $\mathbf{F}^d_4$, which is then concatenated with the skip-connected feature $\mathbf{F}^e_4$ and passed to the next decoder block.
This process continues iteratively until the final decoder output $\mathbf{F}^d_1$ is obtained. Finally, $\mathbf{F}^d_1$ is passed through a Patch Expansion layer to restore spatial resolution and generate the final lesion prediction $\mathbf{Y} \in \mathbb{R}^{1 \times H \times W}$ that matches the input size.

\subsection{Progressive Vision State Space Encoder}

\subsubsection{Convolutional Vision State-Space Block}

\begin{figure}[h]
\centering
\includegraphics[width=\linewidth]{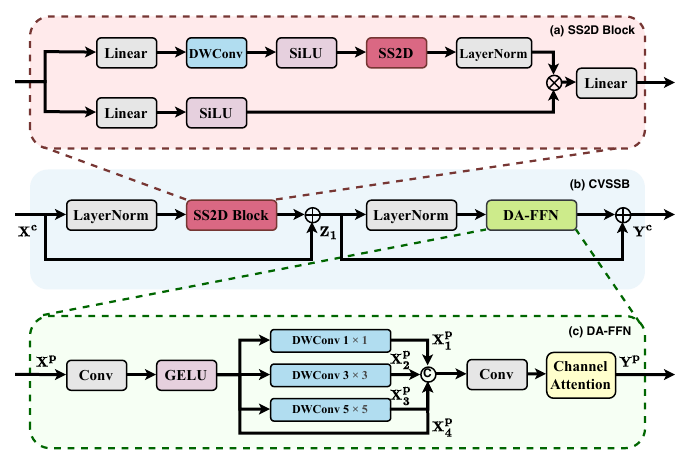}
\caption{The structure of the CVSSB module and its internal components: (a) the SS2D Block, (b) the proposed Convolutional Vision State Space Block (CVSSB), and (c) the proposed Depthwise-Aware FeedForward Network (DA-FFN).}
\label{fig:cvssb_module}
\end{figure}

\begin{figure}[h]
\centering
\includegraphics[width=\linewidth]{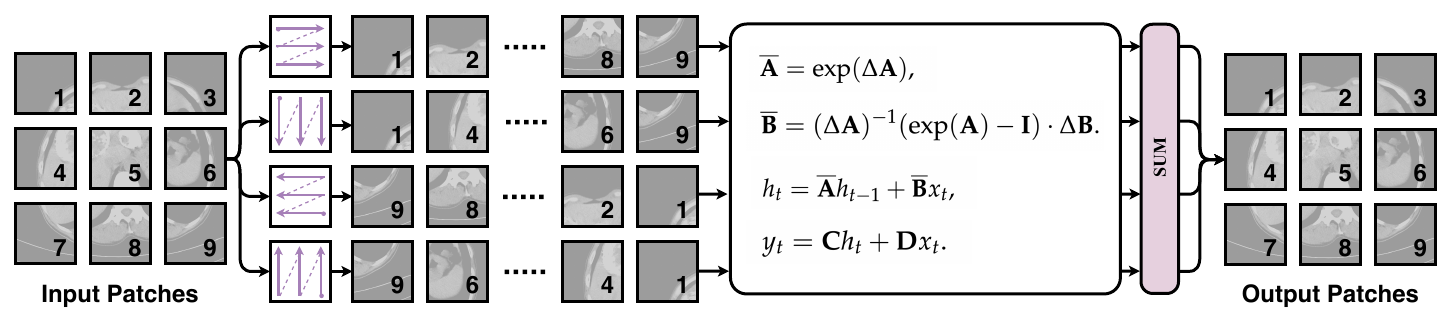}
\vspace{-2em} 
\caption{Illustration of 2D Selective Scan (SS2D). The input feature map is sequentially processed along four different scanning directions, with each directional sequence modeled by a distinct discrete state-space equation to capture long-range dependencies. The directional outputs are then integrated and reshaped to recover the 2D feature map.}

\label{fig:ss2d}
\end{figure}

To simultaneously capture multi-scale fine-grained features and global contextual dependencies while addressing the inherent directional sensitivity of 2D visual data, we propose the CVSSB, as illustrated in Figure.~\ref{fig:cvssb_module}(b). Specifically, we introduce the SS2D Block \cite{vmamba} to model long-range dependencies, as illustrated in Figure.~\ref{fig:cvssb_module}(a). SS2D flattens the image into a one-dimensional sequence, and then applies different discrete state-space equations in parallel to capture long-range dependencies along various directional sequences, as shown in Figure.~\ref{fig:ss2d}. The resulting global representations are then fed into a Depthwise-Aware FeedForward Network (DA-FFN) to further enhance feature interaction and fusion. The processing of the CVSSB can be formulated as follows:

\begin{align}
\begin{aligned}
\mathbf{Z}_1 &= \mathbf{X^c} + \mathrm{SS2DBlock}(\mathrm{LN}(\mathbf{X^c})), \\
\mathbf{Y^c}  &= \mathrm{Conv}(\mathbf{Z}_1) + \mathrm{DA\mbox{-}FFN}(\mathrm{LN}(\mathbf{Z}_1)).
\end{aligned}
\label{eq:cvssb}
\end{align}

Where, $\mathbf{X^c} \in \mathbb{R} ^{P \times H \times W}$ and $\mathbf{Y^c} \in \mathbb{R} ^{\hat{P} \times H \times W}$ denote the input and output features, respectively. 
In the first block of the CVSSB group, $\hat{P}$ is set to $jP$, otherwise, $\hat{P}$ remains equal to $P$. Where $j = 2$ in the PVSS encoder and $j = \frac{1}{2}$ in the HVSS decoder.
$\mathbf{Z_1} \in \mathbb{R} ^{P \times H \times W}$ represents the obtained global representation. $\mathrm{LN}(\cdot)$ denotes Layer Normalization, and $\mathrm{Conv}(\cdot)$ is used to adjust the number of feature channels to filter out redundant features.

\subsubsection{Depthwise-Aware FeedForward Network}

To further enhance the ability to aggregate and represent multi-scale information, we incorporate a DA-FFN, as show in Figure.~\ref{fig:cvssb_module}(c),  which consists of a set of parallel convolutions with varying kernel sizes. These multi-branch convolutions are fused via a channel attention mechanism to emphasize informative feature channels while suppressing redundant ones, ultimately improving the block’s capacity to represent diverse spatial patterns. The definition of DA-FFN can be formulated as follows in Eq.~\ref{eq:daffn}.

\begin{align}
\begin{aligned}
\{
    \mathbf{X}^p_i\}_{i=1}^{4} &= 
        \chi\left(  \sigma\left(
        \mathrm{Conv}\left(\mathbf{X}^p\right) \right) \right),\\
\mathbf{Y}^p &=
\mathrm{CA}\left( 
    \mathrm{Conv} \left( 
        \Phi\left(
            \left\{ 
                \mathrm{DWConv}_{k_i \times k_i}(\mathbf{X}^p_i) 
            \right\},\ 
            \mathrm{Conv}(\mathbf{X}^p_4)
        \right) 
    \right) 
\right).
\end{aligned}
\label{eq:daffn}
\end{align}

Let $\mathbf{X}^p \in \mathbb{R}^{C \times H \times W}$ denote the input feature map. Specifically, the input $\mathbf{X}^p$ is first activated by the SiLU function $\sigma(\cdot)$ and then passed through a convolutional layer to project the channel dimension from $C$ to $4C$. The output is evenly split along the channel dimension by the operation $\chi(\cdot)$ into four chunks. The first three chunks $\mathbf{X}_i^p$ for $i \in {1,2,3}$ are processed by depth-wise convolutions with kernel sizes $k_i \in {1,3,5}$, respectively. The fourth chunk $\mathbf{X}_4^p$ is processed by a standard convolution. These feature maps are then concatenated along the channel dimension via the operation $\Phi(\cdot,\cdot)$, and passed through another convolutional layer that reduces the channel dimension from $4C$ to $2C$. Finally, a channel attention module $\mathrm{CA}$ is applied to highlight informative features and suppress redundant ones, yielding the final output $\mathbf{Y}^p \in \mathbb{R}^{2C \times H \times W}$.

\subsubsection{Haar Wavelet Transform Block}

\begin{figure}[h]
\centering
\includegraphics[width=\linewidth]{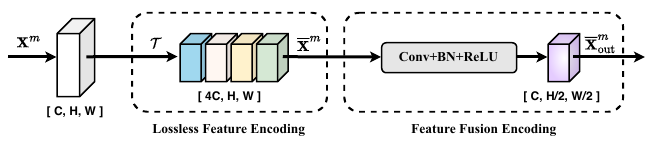}
\caption{The structure of the Haar Wavelet Transform Block.}
\label{fig:hwtb}
\end{figure}

Conventional pooling-based downsampling operations often lead to the loss of spatial information. However, accurate segmentation of HE lesions—characterized by blurred boundaries—requires the preservation of fine-grained features for precise edge delineation. To address this issue, inspired by HWD\cite{hwd,li2022efficient}, we employ a HWTB to achieve lossless downsampling, as shown in Figure.~\ref{fig:hwtb}.

Specifically, the input feature map is denoted as $\mathbf{X}^m \in \mathbb{R}^{C \times H \times W}$. It is first processed by a Haar wavelet transform denoted as $\mathcal{T}$, which decomposes the feature map into four sub-bands:
$\left\{ \mathbf{X}^m_L, \mathbf{X}^m_{HV}, \mathbf{X}^m_{HH}, \mathbf{X}^m_{HD} \right\} \in \mathbb{R}^{C \times \frac{H}{2} \times \frac{W}{2}}$. Where, $\mathbf{X}^m_L$ captures the low-frequency approximation of the original feature, while $\mathbf{X}^m_{HV}$, $\mathbf{X}^m_{HH}$, and $\mathbf{X}^m_{HD}$ retain vertical, horizontal, and diagonal detail information, respectively.
Next, these four sub-band features are concatenated along the channel dimension to obtain $\overline{\mathbf{X}}^m \in \mathbb{R}^{4C \times \frac{H}{2} \times \frac{W}{2}}$.

This effectively compresses the spatial resolution while preserving structural information through channel expansion. To reduce redundancy and restore the channel dimension back to C, a $1 \times 1$ convolution is applied. The result then undergoes batch normalization and ReLU activation, yielding the final output of HWTB, denoted as $\overline{\mathbf{X}}^m_{\text{out}} \in \mathbb{R}^{C \times \frac{H}{2} \times \frac{W}{2}}$.

\subsection{Hybrid Visual State Space Decoder}

To integrate the hierarchical semantic features extracted by the encoder with spatial detail information, we design a Hybrid Visual State Space (HVSS) Decoder, which progressively reconstructs the feature maps to generate the final prediction. As shown in Figure.~\ref{fig:overall}, the decoder consists of four decoder blocks and one Patch Expansion layer.

Each decoding block is responsible for gradually restoring spatial resolution and semantic detail. At each stage, the upsampled decoder feature is fused with the corresponding encoder feature via skip connections to preserve multi-scale contextual information. In the earlier stages (Stage 2 and Stage 3), we adopt a combination of stacked CVSSB modules and upsampling layers. In the deeper stages (Stage 4 and Stage 5), a CBAM module is incorporated to suppress redundant responses and enhance discriminative features.

After processing through the four decoder blocks, the feature map is passed to the Patch Expansion layer, which comprises two consecutive blocks of convolution and bilinear interpolation. This final module outputs the lesion probability map with the same resolution as the input image.

\subsection{Loss Function}

Binary Cross-Entropy (BCE) loss is a widely adopted objective function for binary image classification tasks,
\begin{equation}
\text{Bce}(\mathbf{y},\hat{\mathbf{y}}) =  -\frac{1}{N} \sum_{i=1}^{N} \left[ \mathbf{y}_i \log(\hat{\mathbf{y}}_i) + (1 - \mathbf{y}_i) \log(1 - \hat{\mathbf{y}}_i) \right],
\label{bce}
\end{equation}
where, $N$ denotes the total number of pixels, $\mathbf{y}$ is the ground truth, and $\hat{\mathbf{y}}$ represents the predicted output. Nevertheless, due to the typically small size of lesion regions in medical images, BCE alone often struggles with class imbalance. To address this issue, the Dice loss is incorporated, as it is particularly effective in emphasizing small foreground regions,
\begin{equation}
\text{Dice}(\mathbf{y},\hat{\mathbf{y}}) = 1 - \frac{2 \sum_{i=1}^{N} \mathbf{y}_i \cdot \hat{\mathbf{y}}_i}{\sum_{i=1}^{N} \mathbf{y}_i + \sum_{i=1}^{N} \hat{\mathbf{y}}_i}. \label{dice}
\end{equation}
Consequently, we combine both \eqref{bce} and \eqref{dice},

\begin{equation}
\mathcal{L} =  
\lambda_1 \times \text{Dice}(\mathbf{L}_i,\mathbf{P}_i) + \lambda_2 \times \text{Bce}(\mathbf{L}_i,\mathbf{P}_i)   .
\label{L}
\end{equation}
In our experiments, the weights $\lambda_1$ and $\lambda_2$ are empirically set to 1 by default.

\section{Experiments and Results}

\subsection{Dataset and Implementation Deatils}

\subsubsection{Dataset}

\begin{figure}[h]
\includegraphics[width=\linewidth]{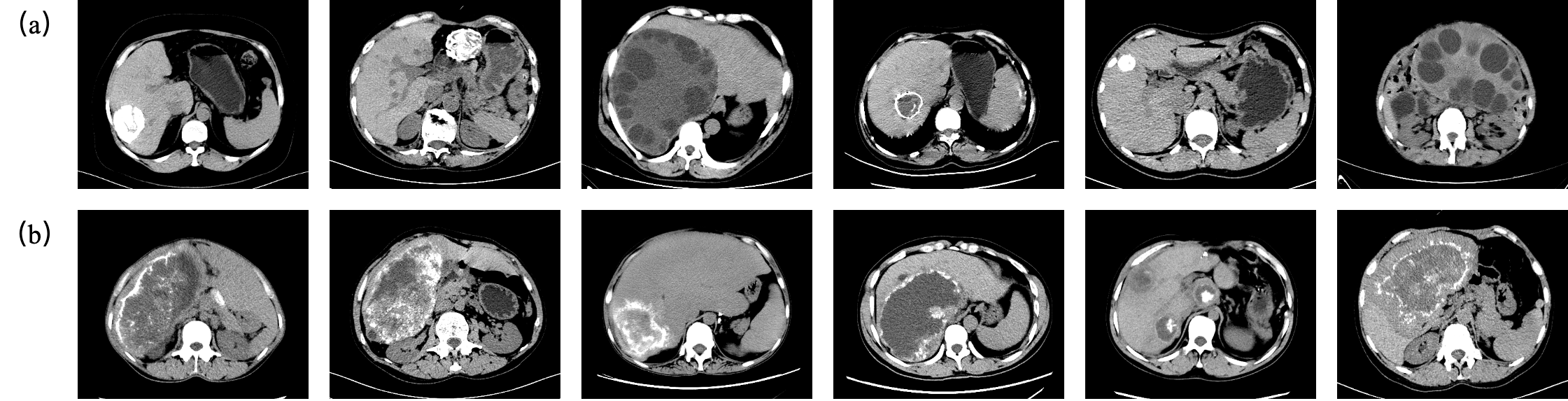}

\caption{Representative examples from the HE dataset. (a) CE; (b) AE.}

\label{fig:dataset}

\end{figure}

We collected abdominal CT scans from 260 patients who were clinically diagnosed with hepatic echinococcosis (HE), including 130 cases of cystic echinococcosis (CE) and 130 cases of alveolar echinococcosis (AE). Representative samples are shown in Figure.~\ref{fig:dataset}.

To protect patient privacy, we removed all metadata that contained personally identifiable information. The original CT Hounsfield Unit (HU) values have a wide dynamic range, which exceeds the perceptual range of the human eye. To address this, we applied a windowing procedure under the guidance of three experienced radiologists. They considered factors such as organ tissues, lesion boundaries, and bone contrast. We set the window width to 150 and the window level to 35. This adjustment mapped the HU values to a standardized range of [-40, 110], enhancing the contrast between different structures. Finally, all CT volumes were converted into 2D slices and split into training, validation, and test sets with a ratio of 80\%, 10\%, and 10\%, respectively.

\subsection{Implementation Details}
Since CT images are single-channel grayscale scans, the input channel number is set to $1$. The output channel is also set to $1$, corresponding to binary segmentation. The patch size in the patch embedding layer is set to $4$. The feature channels in each encoder stage are configured as $[32,\ 64,\ 128,\ 256,\ 512]$, and the number of stacked CVSSB layers in each stage is set to $[2,\ 2,\ 4,\ 2]$. We employed three widely used evaluation metrics in medical lesion segmentation: DSC, Precision, and Recall. These metrics quantitatively assess the overlap and classification performance between the predicted segmentation and the ground truth. 

We trained the model for up to 100 epochs and adopted an early stopping mechanism based on validation loss. Specifically, if the validation loss did not improve over 30 consecutive epochs, the training process was terminated to mitigate overfitting. The AdamW optimizer~\cite{AdamW} was employed with an initial learning rate of 0.001. To adapt the learning rate during training, we used the ReduceLROnPlateau scheduler, which reduce the learning rate by half if no improvement is observed on the validation set for 5 epochs. The batch size was set to 4 in all experiments.

Due to the difficulty of acquiring a large number of CT scans from HE patients, the dataset used in our study was relatively limited. To enhance generalization and reduce overfitting, we applied several data augmentation techniques, including random image rotations, random scaling, random Gaussian smoothing, and random Gaussian noise.

\subsection{Comparison with State-of-the-art Methods}

We conducted extensive comparative experiments on our dataset to quantitatively evaluate the performance of the proposed EAGLE model against several existing medical image segmentation methods. We selected a range of well-established and high-performing baseline models, including UNet\cite{unet}, UNet++\cite{zhou2018unet++}, ResUNet\cite{resunet}, Attention UNet\cite{attentionunet}, TransUNet\cite{transunet}, Swin UNet\cite{swinunet}, HMT-UNet\cite{hmtunet}, and MSVM-UNet\cite{msvmunet}. The results of comparison are shown in Table~\ref{tab:comparison}.

\begin{table}[h]
\centering
\caption{Comparison of different methods on the HE dataset. The optimal values are highlighted in bold, while the second-best values are underlined.}
\label{tab:comparison}
\begin{tabularx}{\textwidth}{CCCC} 
\toprule
\textbf{Methods} & \textbf{DSC (\%) $\uparrow$} & \textbf{Precision (\%)$\uparrow$} & \textbf{Recall (\%)$\uparrow$} \\
\midrule
U-Net\cite{unet} & 85.22 & 85.63 & 86.45 \\
U-Net++\cite{zhou2018unet++} & 85.41 & 85.82 & 86.59 \\
Res-UNet\cite{resunet} & 85.53 & 85.74 & 85.87\\
Attention-UNet\cite{attentionunet} & 85.34 & 85.92 & 86.72 \\
TransUNet\cite{transunet} & 87.62 & 88.09 & 87.66 \\
Swin-UNet\cite{swinunet} & 87.55 & 87.94 & 88.02 \\
HMT-UNet\cite{hmtunet} & 87.72 & \underline{88.41} & 87.84 \\
MSVM-UNet\cite{msvmunet} & \underline{88.15} & 88.39 & \underline{89.26} \\
\midrule
\textbf{EAGLE (Ours)} & \textbf{89.76} & \textbf{88.95} & \textbf{89.47} \\
\bottomrule
\end{tabularx}
\end{table}

Compared to the CNN-based UNet, EAGLE achieves a 4.54\% improvement in Dice Similarity Coefficient (DSC). When compared with the Transformer-based SwinUNet, EAGLE yields a 2.21\% gain in DSC. Additionally, EAGLE outperforms the SSMs-based MSVM-UNet by 1.61\% in DSC. Furthermore, EAGLE also achieves the highest scores in both Precision and Recall metrics. These experimental results demonstrate that EAGLE consistently outperforms existing models in the task of HE lesion segmentation.

\begin{figure}[h]
\centering
\includegraphics[width=\textwidth]{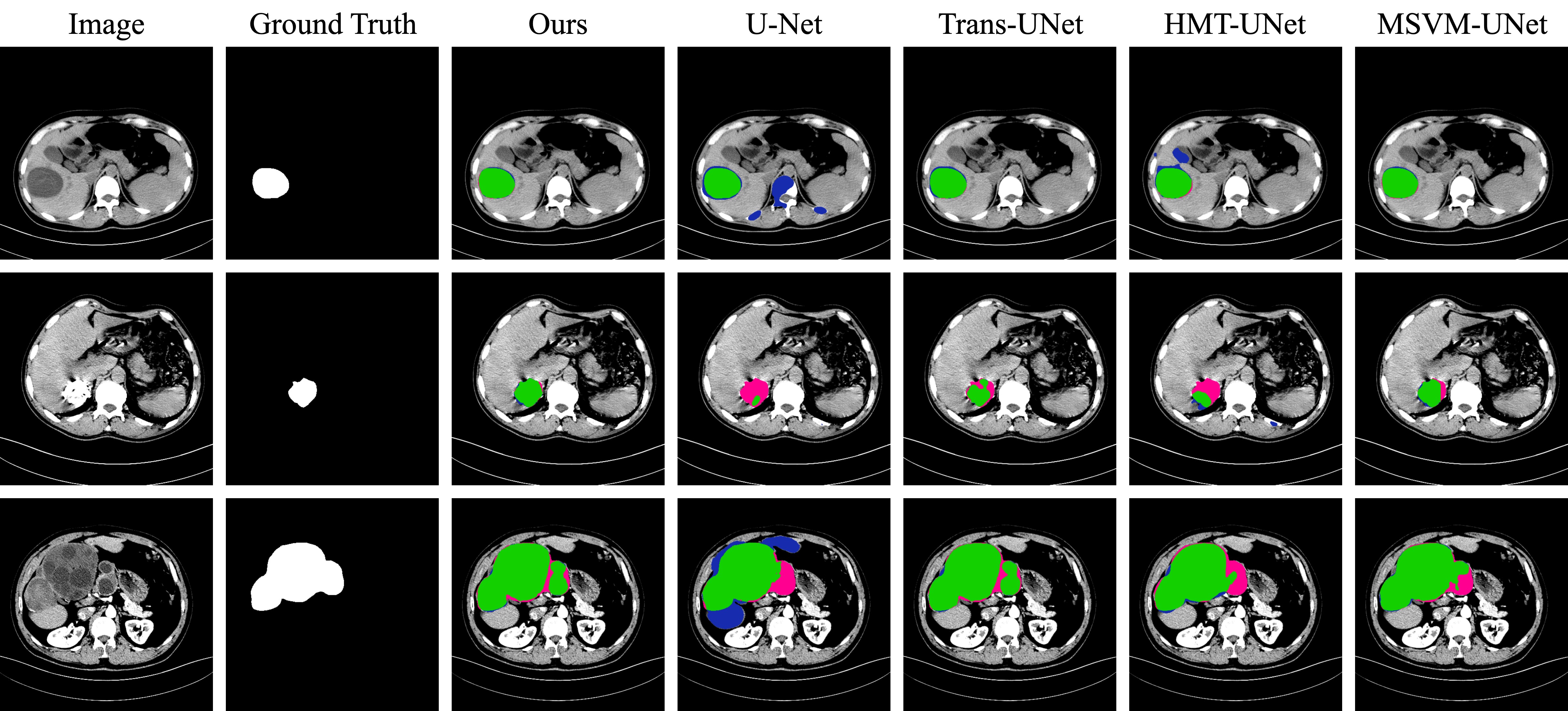}

\caption{Comparison between our proposed EAGLE and other methods on CE lesion segmentation. True positives (TP) are shown in green, false negatives (FN) in hotpink, and false positives (FP) in blue.}
\label{fig:CE_compar}
\end{figure}

\begin{figure}[h]
\centering
\includegraphics[width=\textwidth]{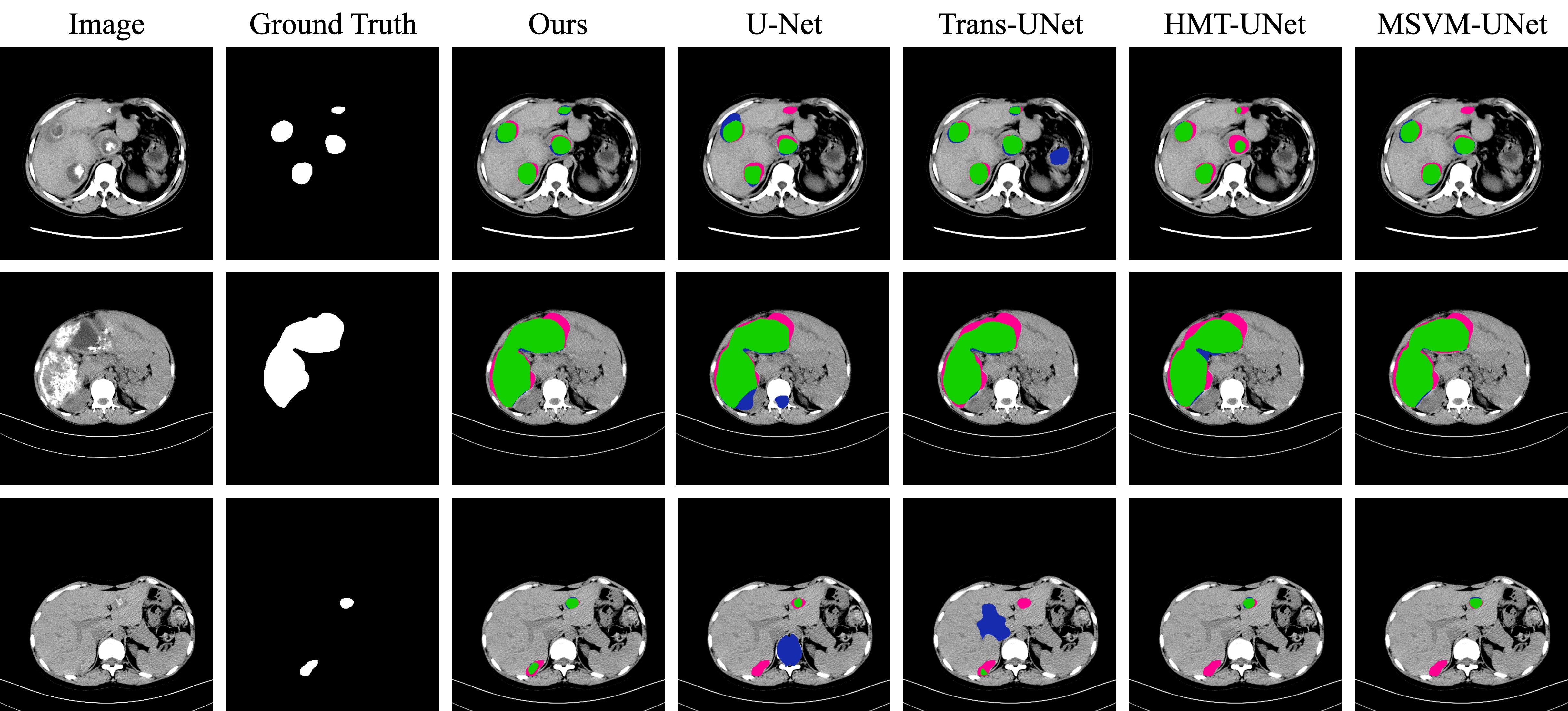}  
\caption{Comparison between our proposed EAGLE and other methods on AE lesion segmentation. True positives (TP) are shown in green, false negatives (FN) in hotpink, and false positives (FP) in blue.}
\label{fig:AE_compar}
\end{figure}

The comparative segmentation results of different methods on CE and AE are illustrated in Figure.\ref{fig:CE_compar} and Figure.\ref{fig:AE_compar}. For the pure CNN-based method UNet, its limited ability to model global contextual information prevents it from accurately capturing the spatial relationships among lesions, liver, bones, and other internal organs. As a result, it tends to misclassify spinal regions with similar intensities as calcified lesions, leading to suboptimal segmentation performance. Although the Transformer-based TransUNet can model global dependencies, its effectiveness is constrained by the limited size of our dataset. Since Transformer architectures typically require large-scale datasets to achieve strong generalization, TransUNet performs poorly in this domain. As for the SSM-based MSVM-UNet, while it demonstrates good performance in segmenting the main lesion regions, it still struggles with lesions exhibiting blurred boundaries and high invasiveness, particularly those of the AE type, resulting in segmentation inaccuracies. 

EAGLE leverages the collaborative design of the proposed CVSSB and HWTB modules to effectively integrate local and global features while filtering out redundant information. Additionally, during the downsampling stage, spatial information is elegantly compressed into the channel dimension, mitigating feature loss. As a result, the model maintains superior segmentation performance under challenging conditions such as low contrast, noise interference, and blurred boundaries.

\subsection{Ablation Analysis}

\subsubsection{CVSSB Stacking Depth}
To evaluate the impact of the number of stacked CVSSB at different stages on the model’s performance, we conducted an ablation study. Specifically, we varied the value of $L$, which denotes the number of CVSSB layers in each stage, and observed its effect on segmentation accuracy. The detailed results are presented in Table~\ref{tab:cvssb_ablation}.

\begin{table}[h]
\centering
\caption{Ablation study on different CVSSB stacking configurations. The optimal values are highlighted in bold. \( L_i \) denotes the number of CVSSB blocks in the \( i \)-th stage. Configurations marked with \texttt{CUDA Out of Memory} could not be evaluated due to GPU memory limitations.}
\label{tab:cvssb_ablation}
\begin{tabularx}{\textwidth}{>{\centering\arraybackslash}X 
                        >{\centering\arraybackslash}p{2.4cm} 
                        >{\centering\arraybackslash}p{2.7cm} 
                        >{\centering\arraybackslash}p{2.4cm}}
\toprule
\textbf{CVSSB Layers(\(L_i\))} & \textbf{DSC (\%)} & \textbf{Precision (\%)} & \textbf{Recall (\%)} \\
\midrule
w/o CVSSB                       & 87.15 & 87.40 & 86.90 \\
$[1,\ 1,\ 1,\ 1]$             & 88.09 & 88.10 & 87.95 \\
$[2,\ 2,\ 2,\ 2]$             & 88.65 & 88.51 & 88.42 \\
\rowcolor{gray!20}
$[3,\ 3,\ 3,\ 3]$             & \multicolumn{3}{c}{\texttt{CUDA Out of Memory}} \\
$[2,\ 2,\ 2,\ 3]$             & 88.74 & 89.02 & 88.69 \\
$[2,\ 2,\ 3,\ 2]$             & 89.11 & 89.32 & 88.92 \\
\textbf{\boldmath$[2,\ 2,\ 4,\ 2]$} & \textbf{89.76} & \textbf{90.12} & \textbf{89.45} \\
\rowcolor{gray!20}
$[2,\ 2,\ 5,\ 2]$             & \multicolumn{3}{c}{\texttt{CUDA Out of Memory}} \\
\bottomrule
\end{tabularx}
\end{table}

We conducted eight ablation experiments to investigate the effect of different CVSSB configurations. Starting with a baseline model where standard convolutions replaced CVSSB, we gradually increased the number of stacked layers. Setting all stages to have three CVSSB led to out-of-memory errors, making training infeasible. Through extensive testing, we observed that increasing the number of CVSSB blocks in Stage 3 yielded more significant performance gains than in other stages. Based on this insight, we focused on enhancing Stage 3 while keeping the overall parameter count within a manageable range. The final configuration, $L = [2, 2, 4, 2]$, proved to be the optimal setting.

\subsubsection{Proposed Modules}
To evaluate the individual contributions of the proposed CVSSB, DA-FFN, and HWTB modules to the overall performance of our model, we conducted a series of ablation experiments using Vanilla UNet as the baseline. Since DA-FFN is an integral component of CVSSB, introducing DA-FFN requires CVSSB to be present. It is worth noting that the number of CVSSB blocks in each stage is set to $L = [2, 2, 4, 2]$, which has been shown to be the optimal setting in previous experiments.

The detailed results are summarized in Table~\ref{tab:ablation}. When all three modules are applied together, the model achieves improvements of 3.1\% in DSC, 1.68\% in Precision, and 2.24\% in Recall compared to the baseline. Specifically, the proposed CVSSB is designed to better aggregate local and global features, and incorporating CVSSB alone leads to a 1.34\% increase in DSC over the Vanilla UNet. Building on this, DA-FFN further enhances multi-scale representation capabilities, contributing an additional 2\% improvement in DSC when used in conjunction with CVSSB. Finally, HWTB mitigates the loss of fine-grained details during the downsampling process and helps preserve valuable information, thereby improving segmentation accuracy. These results demonstrate that the proposed CVSSB, DA-FFN, and HWTB modules effectively enhance the model’s segmentation performance and validate the effectiveness of our architectural design.

\begin{table}[H]
\centering
\caption{Ablation study of different modules. The optimal values are highlighted in bold.}
\label{tab:ablation}
\begin{tabularx}{\textwidth}{ccc|>{\centering\arraybackslash}X >{\centering\arraybackslash}X >{\centering\arraybackslash}X}
  \toprule
  \textbf{CVSSB} & \textbf{DA-FFN} & \textbf{HWTB} & \textbf{DSC (\%)} & \textbf{Precision (\%)} & \textbf{Recall (\%)} \\
  \midrule
  $\times$     & $\times$     & $\times$     & 85.22 & 85.63 & 86.45 \\
  \midrule
  \checkmark   & $\times$     & $\times$     & 86.56 & 86.11 & 86.71 \\
  \checkmark   & \checkmark   & $\times$     & 87.22 & 86.67 & 87.21 \\
  $\times$     & $\times$     & \checkmark   & 85.75 & 86.02 & 86.63 \\
  \midrule
  \checkmark   & $\times$     & \checkmark   & 87.23 & 86.60 & 87.25 \\
  \midrule
  \checkmark   & \checkmark   & \checkmark   & \textbf{88.32} & \textbf{87.31} & \textbf{88.69} \\
  \bottomrule
\end{tabularx}
\end{table}

\section{Conclusion}

To further enhance segmentation performance, we introduce two novel modules. The CVSSB combines the strengths of convolution and state space modeling to facilitate comprehensive feature fusion. The HWTB performs lossless spatial downsampling using Haar wavelet transform, thus preserving fine-grained details that are often lost during the encoding stage. We conduct extensive experiments on a self-collected dataset comprising 260 HE patients. EAGLE achieves excellent performance with a DSC of 89.76\%, surpassing the previous best method MSVM-UNet by 1.61\%. Through thorough comparative and ablation studies, we validate the effectiveness of EAGLE and its key components in improving segmentation accuracy. In summary, EAGLE demonstrates strong potential in HE lesion segmentation and provides a valuable tool to assist clinicians in more efficient and accurate diagnosis of hepatic echinococcosis.

\bibliographystyle{splncs04}
\bibliography{references}
\end{document}